\documentclass[%
aps,preprint,prc,amsmath,amssymb,nofootinbib,superscriptaddress%
]{revtex4}

\usepackage[utf8]{inputenc}
\usepackage[usenames]{color}
\usepackage[dvipsnames]{xcolor}
\usepackage{graphicx}
\usepackage{amsmath}
\usepackage{amsfonts}
\usepackage{amssymb}
\usepackage{mathrsfs}
\usepackage{bm}
\usepackage{verbatim}
\usepackage{cancel}
\usepackage{comment}
\usepackage[normalem]{ulem}
\usepackage{float}
\usepackage{textcomp}
\usepackage{booktabs}
\usepackage{braket}

\usepackage{hyperref}
\hypersetup{
  colorlinks=true, 
  linkcolor=ForestGreen,
}

\usepackage{multirow,makecell}

\newcommand{\chn}[3]{{{}^{#1}\!{#2}_{#3}}}
\newcommand{\cs}[2]{\chn{#1}{S}{#2}}
\newcommand{\chp}[2]{\chn{#1}{P}{#2}}
\newcommand{\cd}[2]{\chn{#1}{D}{#2}}

\newcommand{\csd}{{\cs{3}{1}-\cd{3}{1}}}

\newcommand{\VSO}[1]{{\widehat{\mathcal{O}}^{(#1)}}}
\newcommand{\NNLO}{N$^2$LO}

\graphicspath{{./figs/}}

\begin{document}

\title{Contact operators in renormalization of attractive singular potentials}

\author{Rui Peng}
\affiliation{School of Physics, Peking University, Beijing 100871, China}

\author{Bingwei Long}
\email{bingwei@scu.edu}
\affiliation{College of Physics, Sichuan University, Chengdu 610065, China}
\affiliation{Southern Center for Nuclear-Science Theory (SCNT), Institute of Modern Physics, Chinese Academy of Sciences, Huizhou 516000, Guangdong Province, China}

\author{Fu-Rong Xu}
\email{frxu@pku.edu.cn}
\affiliation{School of Physics, Peking University, Beijing 100871, China}
\affiliation{Southern Center for Nuclear-Science Theory (SCNT), Institute of Modern Physics, Chinese Academy of Sciences, Huizhou 516000, Guangdong Province, China}

\date{October 3, 2024}

\begin{abstract}

We discuss renormalization of chiral nuclear forces in the $\chp{3}{0}$ channel of $NN$ scattering at next-to-next-to leading order (\NNLO) if the one-pion exchange is treated nonperturbatively at leading order. The matrix elements of the subleading contact potentials become nearly dependent of each other for the so-called exceptional ultraviolet momentum cutoff, making it difficult to determine the strengths of those contact potentials from the empirical phase shifts, as reported in Ref.~\cite{Gasparyan:2022isg}. We argue that this issue can be resolved by adjusting the strategy by which the low-energy constants are deduced from the data, thus making those exceptional cutoffs amenable to chiral effective field theory.

\end{abstract} 
\maketitle

\section{Introduction\label{sec:intro}}

Renormalization-group (RG) invariance plays an important role in constructing nucleon-nucleon ($NN$) forces in the framework of chiral effective field theory (ChEFT). Even though the chiral potentials are regularized by an arbitrarily chosen ultraviolet momentum cutoff $\Lambda$, one nonetheless wishes the observables calculated with these regularized interactions come out to be independent of $\Lambda$, hence free from model-dependent assumptions about the underlying physics associated with $\Lambda$. In order to achieve cutoff independence, one may need to adjust the set of short-range forces, which appear in the form of $NN$ contact interactions in the two-body sector, so that the $\Lambda$ dependence, if any, could be absorbed by these contact interactions, or so-called counterterms~\cite{Kaplan:1996xu, Nogga:2005hy, Birse:2005um, Long:2011xw, Long:2012ve, PavonValderrama:2016lqn, Bira:2020rv}.
In this context, RG invariance is equivalent to cutoff independence, and organization of counterterms, or power counting of counterterms, is constrained by the requirement of RG invariance. In ChEFT-based studies of nuclear electroweak currents~\cite{PavonValderrama:2014zeq, Shi:2022blm, Liu:2022cfd} and fundamental symmetries~\cite{deVries:2020loy, Cirigliano:2018hja, Cirigliano:2020dmx}, RG invariance has also proven a valuable guideline. We focus in the present paper on renormalization of subleading interactions once one-pion exchange (OPE), the most important long-range chiral nuclear force, is renormalized at leading order (LO). More specifically, we address the issue raised in a recent work~\cite{Gasparyan:2022isg} which states that the procedure to determine the low-energy constants (LECs) of subleading counterterms does not lead to RG-invariant EFT amplitudes.

Renormalization of the tensor force of OPE is the challenging because it is singular and attractive in various triplet channels, such as $\csd$, $\chp{3}{0}$, and so on, that is, behaving like $-1/r^3$ at short distances in these partial waves~\cite{RevModPhys.43.36}. In each of those partial waves, a short-range counterterm must be added to the LO potentials, otherwise the singular attraction can overcome the kinetic energy and the centrifugal barrier and collapse the $NN$ system. Deficiency of short-range interactions of this sort can be detected by inspecting RG invariance of the EFT amplitude~\cite{Beane:2000wh, Long:2007vp, Hammer:2011kg, Odell:2019wjq}.

Much of our attention will be on renormalization at higher orders once the tensor force of OPE is renormalized at LO. To disentangle increasingly complicated structures of higher-order forces, including long-range forces like two-pion exchanges (TPEs), treating higher-order terms in perturbation theory was advocated, akin to the well-known distorted-wave expansion~\cite{Long:2007vp, Long:2011qx, Long:2012ve, PavonValderrama:2016lqn}. 
However, Ref.~\cite{Gasparyan:2022isg} found recently that around certain values of $\Lambda$, referred to as the ``genuine exceptional cutoffs'', the otherwise independent subleading contact operators in $\chp{3}{0}$ become dangerously correlated; therefore, they can not be determined in the particular way laid out in Ref.~\cite{Long:2011qx}.
Similar numerical difficulty to determine LECs was also reported in Ref.~\cite{Shi:2022blm} for $\csd$. If one has to exclude some cutoff values in implementing the EFT, RG invariance will appear to be violated.
But it will be incredible if RG invariance as a general guideline of EFTs fails in a particular theory. It is more likely that the particular procedure of determining the values of the LECs used in Refs.~\cite{Long:2007vp, Long:2011qx} leads to the dilemma. This is the direction explored in the paper. To avoid complication due to the coupled-channel dynamics, we study the uncoupled channel of $\chp{3}{0}$ in the paper.

In order to resolve the issue it is useful to reduce the unwelcome correlation between the subleading counterterms. This takes us back to renormalization at LO~\cite{Beane:2000wh, Beane:2001bc}. Because of the approximate nature of any EFTs, we have some leeway in setting the conditions under which the LECs are determined, to the extent allowed by the power counting. Our strategy is to exploit this freedom. If such conditions at LO are let change slightly near those exceptional cutoffs, a cutoff variation of the LO amplitude will appear but it is no worse than the expected subleading correction. More importantly, the subleading counterterms in this set up will regain the capacity to control \emph{independently} the higher-order amplitude, thus compensating the cutoff variation intentionally delegated from the LO. The end result is that the EFT amplitude at each order has only a cutoff variation in accordance with the power counting.

In Sec.~\ref{sec:LO}, the structure of the LO $\chp{3}{0}$ amplitude is analyzed. The mechanism by which the correlation between the subleading contact operators arises is examined in Sec.~\ref{sec:Correlation}, and an alternative way to determine the LO LEC is proposed there. We then show in Sec.~\ref{sec:NNLO} how this alternative works to ensure the RG invariance at next-to-next-to leading order ({\NNLO}), followed by a summary in Sec.~\ref{sec:summary}.

\section{Leading order\label{sec:LO}}

To obtain the LO off-shell $NN$ $T$-matrix in $\chp{3}{0}$, we use the partial-wave Lippmann-Schwinger (LS) equation to iterate the LO potential $V^{(0)}$ nonperturbatively:
\begin{equation}
    T^{(0)}(p',p; k)=V^{(0)}(p',p)+\int^\Lambda_0 dl\, l^2\, V^{(0)}(p', l) G_0(l; k) T^{(0)}(l, p; k) \, ,
    \label{equ:LSElo}
\end{equation}
where $G_0(l; k) \equiv (k^2 - l^2 +i\epsilon)^{-1}$ is the nonrelativistic propagator of free $NN$ states, $\Lambda$ the momentum cutoff, $k$ the center-of-mass (CM) momentum, $p$ ($p'$) the incoming (outgoing) relative momentum. Here we have let the potentials absorb the factor of the nucleon mass $m_N$. The LS integral equation above is often schematically written as
\begin{equation}
    T^{(0)} = V^{(0)} + V^{(0)} G_0 T^{(0)} \, .
\end{equation}
The on-shell amplitude $T^{(0)}(k) \equiv T^{(0)}(k, k; k)$ is related to the phase shift $\delta^{(0)}$ by
\begin{equation}
    T^{(0)}(k) = \frac{i}{k\pi}\left[e^{2i\delta^{(0)}} - 1 \right]\, .    
\end{equation}

In $\chp{3}{0}$ where our interest lies $V^{(0)}$ consists of the long-range force of OPE, $V_\pi$, and a short-range potential $V_S^{(0)}$. Partial-wave decomposition of long-range forces is nontrivial but can be found in the literature, e.g., in Refs.~\cite{Kaiser:1997mw, Epelbaum:1999dj}. There are more than one way to write contact potentials from the chiral Lagrangian~\cite{Bira:1996, Kaplan:1996xu, EPELBAOUM1999413, Long:2012ve}. We follow Refs.~\cite{Long:2012ve, Long:2011xw, Long:2011qx, Peng_2020} to write all the contact forces as combinations of separable terms that usually have straightforward partial-wave decomposition. For instance, 
\begin{equation}
    V_S^{(0)} =  C^{(0)} \VSO{0} \, ,
\end{equation}
with $\widehat{\mathcal{O}}^{(0)}$ defined as
\begin{equation}
    \langle p'\; \chp{3}{0} | \VSO{0} | p\; \chp{3}{0} \rangle = p' p \, . 
\end{equation}
The superscript ``(0)'' of $C^{(0)}$ indicates that this is the LO value of $C$, and it could receive corrections at higher orders:
\begin{equation}
    C = C^{(0)} + C^{(1)} + C^{(2)} + \cdots
\end{equation}

\begin{figure}
    \centering
    \includegraphics[scale=0.8]{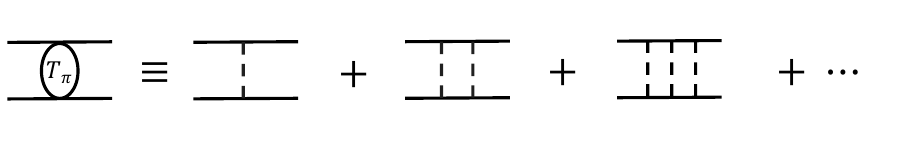}
    \caption{Diagrammatic breakdown of $T_\pi$. The dashed and solid lines represent the pion and nucleon propagators, respectively.}
    \label{fig:FeynTpi}
\end{figure}

It will prove useful to reformulate $T^{(0)}$ using the two-potential trick~\cite{Kaplan:1996xu, Long:2012ve, Gasparyan:2022isg}. Defining $T_\pi$ as the resummation of OPE,
\begin{equation}
    T_\pi(p',p; k) = V_\pi(p',p) + 
    \int^\Lambda_0 dl\, l^2\, V_\pi(p', l) G_0(l; k) T_\pi(l, p; k) \, ,
    \label{eqn:Tpi_Vpi}
\end{equation}
as also diagrammatically illustrated in Fig.~\ref{fig:FeynTpi}, we write $T^{(0)}$ as
\begin{equation}
    T^{(0)}(p',p; k) = T_\pi(p',p; k) + \frac{\chi_\pi(p'; k)\chi_\pi(p; k)}{1/C^{(0)} - I(k)} \, , 
    \label{eqn:TLOWChipiIk}
\end{equation}
where $\chi_\pi(p; k)$ and $I(k)$ are defined as follows: 
\begin{align}
    \chi_\pi(p; k) &\equiv p +  \int^\Lambda_0 dl\, l^3 G_0(l; k) T_\pi(l, p; k) \, , \label{eqn:DefChiPik} \\ 
    I(k) &\equiv \int^\Lambda_0 dl\, l^3 G_0(l; k) \chi_\pi(l; k) \, .
    \label{eqn:DefIk}
\end{align}    
as will be discussed below.

One of interesting properties of OPE is that its attractive tensor force will create an increasing number of deep, spurious bound states, as the cutoff $\Lambda$ grows, in addition to the physical bound state of the deuteron, regardless of the angular momentum $L$. Since their momenta are outside the validity range of ChEFT, these spurious states do not change low-energy observables interested in the ChEFT applications~\cite{Nogga:2005hy}, but they have impacts on off-shell quantities. For instance, the RG running $C^{(0)}(\Lambda)$ has been shown to diverge at some critical values of $\Lambda$, denoted by $\Lambda_\star$ in the paper, when separable cutoff regularizations are used~\cite{Nogga:2005hy, Long:2011qx, Long:2011xw}, a phenomenon known as limit cycles~\cite{Beane:2000wh}. Equation~\eqref{eqn:TLOWChipiIk} can be useful in solving for the value of $C^{(0)}$ near $\Lambda_\star$ where it becomes large.

\begin{figure}
    \centering
    \includegraphics[width = 0.7\textwidth]{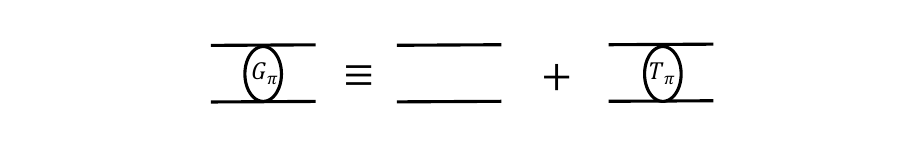}
    \includegraphics[width = 0.8\textwidth]{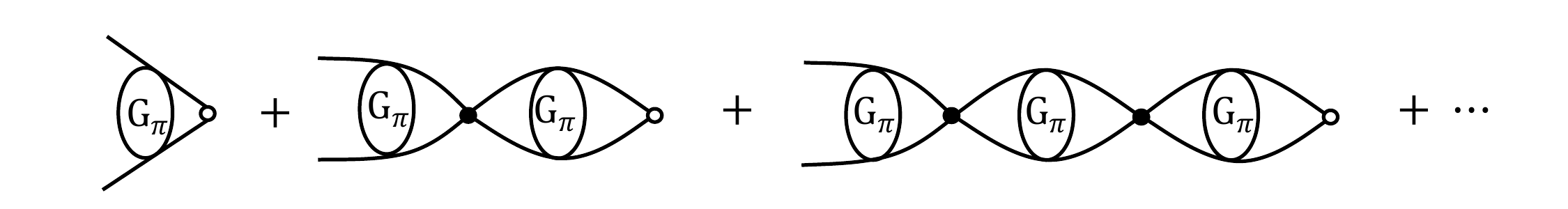}
    \caption{
    In the second line are the Feynman diagrams for $\chi(k)$ while the first line defines $G_\pi$. The solid dot represents the LO contact interaction $V_S^{(0)}$. The circle represents a special vertex function that depends only on the incoming relative momentum $v(p', p) = p$. For definitions of other symbols, see Fig.~\ref{fig:FeynTpi}. 
    }
    \label{fig:FeynPsiLO}
\end{figure}

The divergence of $C^{(0)}(\Lambda)$ has interesting connection with the short-distance structure of the coordinate-space wave function. Since the $P$-wave wave functions $\psi(r; k)$ always vanish at the origin, we study instead their radial derivative at $r = 0$ which turns out to be quite relevant for matrix elements of {\NNLO} contact potentials. $\psi(r; k)$ can be written in terms of $T^{(0)}$ as
\begin{equation}
    \psi(r; k) = \langle k\; \chp{3}{0} | 1 + G_0 T^{(0)} | r\; \chp{3}{0} \rangle \, .
\end{equation}
Up to a numerical factor, we can relate its radial derivative to the following expression:
\begin{equation}
      \chi(k; \Lambda) \equiv k + \int^\Lambda_0 dl\, l^3 G_0(l; k) T^{(0)}(l, k; k) \propto \frac{d \psi}{dr}\Big{|}_{r = 0} \, .
\end{equation}
Using Eq.~\eqref{eqn:TLOWChipiIk}, one finds that $\chi(k; \Lambda)$ is in fact the sum of a geometric series of Feynman diagrams, as shown in Fig.~\ref{fig:FeynPsiLO}. The geometric series are readily resummed in terms of $\chi_\pi(k; k)$ and $ I(k)$:
\begin{equation}
   \chi(k; \Lambda)  = \frac{\chi_\pi(k; k)}{1 - C^{(0)} I(k)} \, . \label{eqn:DefChiK}
\end{equation}
This equation suggests that at $\Lambda_\star$ where $C^{(0)}$ diverges $\chi(k; \Lambda)$ approaches zero regardless of the value of $k$. Therefore, $\chi(k; \Lambda)$ is expected to have the following form near $\Lambda_\star$:
\begin{align}
    \chi(k; \Lambda) &= \left(\Lambda - \Lambda_\star\right)^\alpha \frac{\chi_\pi(k)}{I(k)} \, , \label{eqn:ChiKLambdaStar}
\end{align}
where $\alpha > 0$.

For later use, we introduce a quantity similarly defined that is proportional to the third radial derivative of $\psi(r; k)$:
\begin{equation}
  \phi(k; \Lambda) \equiv k^3 + \int^\Lambda_0 dl\, l^5 G_0(l; k) T^{(0)}(l, k; k) \propto \frac{d^3 \psi}{dr^3}\Big{|}_{r = 0} \, .
\end{equation}
Although more complicated, $\phi(k)$ can also be written in terms of $\chi_\pi(k; k)$ and $I(k)$:
\begin{equation}
   \phi(k; \Lambda)  = \phi_\pi(k; k) + \frac{\chi_\pi(k; k)I_2(k)}{1/C^{(0)} - I(k)} \, , \label{eqn:DefChi2K}
\end{equation}
where
\begin{align}
    \phi_\pi(p; k) &\equiv p^3 + \int^\Lambda_0 dl\, l^5 G_0(l; k) T_\pi(l, p; k) \, , \label{eqn:DefChi2Pik} \\
    I_2(k) &\equiv \int^\Lambda_0 dl\, l^3 G_0(l; k) \phi_\pi(l; k) \, .
    \label{eqn:DefI2k}
\end{align}
Unlike $\chi(k; \Lambda_\star)$, $\phi(k; \Lambda_\star)$ does not vanish, as also numerically verified by us.

\begin{figure}
    \centering
    \includegraphics[width = 0.5\textwidth]{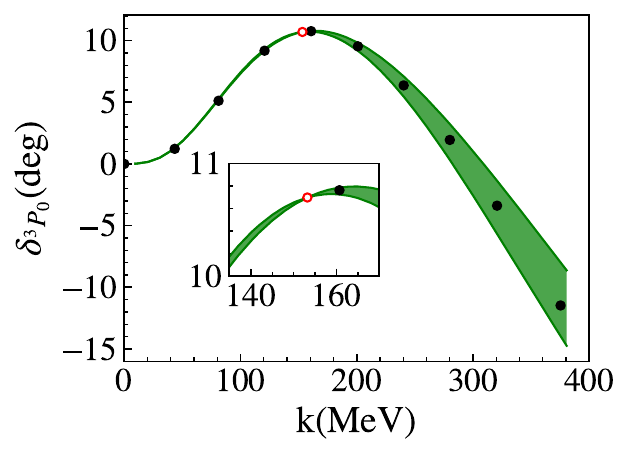}
    \caption{The LO $\chp{3}{0}$ phase shifts as a function of $k$. The solid circles are the PWA phase shifts, and in particular the red empty circle is the input to determine $C^{(0)}$. The band is produced by varying $\Lambda$ from $800$ MeV to $3200$ MeV.}
    \label{fig:3p0_LO}
\end{figure}

The value of $C^{(0)}$ at a given $\Lambda$ is determined using a particular value of the $\chp{3}{0}$ phase shifts at $k = \mu_0$ as the input, $\delta_\text{In}(\mu_0)$. We use $\mu_0 = 153.216$ MeV throughout this paper. $\delta_\text{In}$ is conventionally chosen to be a fixed value for all cutoffs. Therefore, cutoff independence of $T^{(0)}(k)$ will be perfectly satisfied at $k = \mu_0$ by construction. But residual cutoff dependence remains for all other $k$'s in the ChEFT domain, with the expectation that it is of higher order:
\begin{equation}
    T^{(0)}(k; \Lambda) = T^{(0)}(k; \infty)\left[ 1 + \mathcal{O}\left( \frac{k}{\Lambda}\right) \right] \, .
\end{equation}
This is demonstrated in Fig.~\ref{fig:3p0_LO} where we choose $\delta_\text{In}$ to be the empirical value from the partial-wave analyses (PWA) of the Nijmegen group~\cite{nn.online, Stoks:1993tb}, and the cutoff variation is represented by a band. Since perfect RG invariance is not enjoyed at most values of $k$, it is not necessary to be enjoyed at $k = \mu_0$, another arbitrarily chosen momentum scale. In fact, we will intentionally let $\delta_\text{In}$ take a ``wrong'' value, i.e., be off the PWA value by a small amount that is comparable with a subleading correction, in carefully chosen windows of $\Lambda$. This is to remove an undesired correlation between two of the {\NNLO} contact operators, the topic of next section.

\section{Correlation between {\NNLO} contact operators\label{sec:Correlation}}

The NLO potential---the $\mathcal{O}(Q)$ correction to LO--- vanishes in $\chp{3}{0}$ and nontrivial corrections start at {\NNLO}---the $\mathcal{O}(Q^2)$ correction to LO. Our notation of orders is different from, e.g., Refs.~\cite{Epelbaum:1999dj, Entem:2002sf} where NLO ({\NNLO}) labels the $O(Q^2)$ ($O(Q^3)$) correction to the LO potential. The long-range part of the {\NNLO} potential $V^{(2)}$ is the TPE potential composed of vertexes with chiral index $\nu = 0$, $V_{2\pi}^{(0)}$. 

The {\NNLO} short-range potential $V_S^{(2)}$ has a new operator with fourth-degree powers in momenta~\cite{Long:2011qx, Long:2011xw}: $D \VSO{2}$ where $\VSO{2}$ is defined as
\begin{equation}
    \langle p'\; \chp{3}{0} | \VSO{2} | p\; \chp{3}{0} \rangle = p' p \left({p'}^2 + p^2 \right)\, .
\end{equation}
With a naive dimensional estimate $D/C \sim \Lambda^{-2}$, $D \VSO{2}$ does not necessarily appear two orders smaller than the LO contact term $C \VSO{0}$, because the intermediate states have momenta up to $p (p') \simeq \Lambda$. An implementation of the subleading interaction in which $D \VSO{2}$ is reliably restrained from making a larger impact than $C \VSO{0}$ is to treat {\NNLO} potentials in perturbation theory. First, we let $C$ be corrected at {\NNLO} so that $V_S^{(2)}$ has two terms
\begin{equation}
    V_S^{(2)} = C^{(2)} \VSO{0} + D^{(0)} \VSO{2}\, .
\end{equation}
Second, we treat $V_S^{(2)}$ as a perturbation on top of the nonperturbative LO, in a fashion similar to the distorted-wave expansion. By adjusting $C^{(2)}$, we are able to keep 
the whole of $V_S^{(2)}$ small while leaving $C^{(0)}$ intact at LO. Advocates of the perturbative formulation of subleading interactions for nuclear EFTs can be found in, e.g., Refs.~\cite{Kaplan:1996xu, vanKolck:1998bw, Long:2007vp}. It is helpful to take stock and to list the $\chp{3}{0}$ potentials order by order up to {\NNLO}:
\begin{equation}
\begin{aligned}
    V^{(0)} =& V_\pi + V_S^{(0)} \, , \\
    V^{(1)} =& 0 \, , \\
    V^{(2)} =& V_{2\pi}^{(0)} + V_S^{(2)} \, ,
\end{aligned}
\end{equation}
with the LO potential $V^{(0)}$ understood to be iterated nonperturbatively.

Cutoff independence of the subleading amplitudes was demonstrated numerically in Refs.~\cite{Long:2007vp, Valderrama:2009ei, PavonValderrama:2011fcz, Long:2011qx, Long:2011xw, Long:2012ve} for the values of $\Lambda$ sampled therein. With $V^{(2)}$ treated as perturbation, the {\NNLO} correction to the on-shell partial-wave amplitude $T^{(2)}$ is given by
\begin{equation}
    T^{(2)} = (1 + T^{(0)} G_0) V^{(2)} (1 + G_0 T^{(0)}) \, ,
    \label{equ:T2_pert}
\end{equation}
or equivalently, the first term in the distorted-wave expansion:
\begin{equation}
    T^{(2)}(k) = \langle \psi^-_k \; \chp{3}{0} | V^{(2)} | \psi^+_k\; \chp{3}{0} \rangle \, ,
\end{equation}
where $\psi^{+}_k$ ($\psi^{-}_k$) is the in-state (out-state) scattering state generated by the LO potential.
The unitarity of the $S$-matrix requires $T^{(2)}$ be related to the {\NNLO} phase shift by
\begin{equation}
    \delta^{(2)}(k) =  - \frac{\pi}{2} k e^{-2i\delta^{(0)}}T^{(2)}(k) \, .
    \label{equ:delta_rho}
\end{equation}

However, if the procedure prescribed in Ref.~\cite{Long:2007vp, Long:2011qx, Long:2011xw} is followed exactly, meticulous scan of the cutoff value will show that within certain narrow windows of $\Lambda$, cutoff independence of the {\NNLO} $T$ matrix breaks down, as pointed out in Ref.~\cite{Gasparyan:2022isg}. A closer investigation reveals that the sensitivity to $\Lambda$ under these peculiar circumstances is caused by strong correlation between $\VSO{0}$ and $\VSO{2}$ when they are sandwiched between the LO scattering states $\psi_k^\pm$, even though their matrix elements between plane waves are obviously independent. Because the matrix elements of $\VSO{0}$ and $\VSO{2}$ become approximately dependent, one can not determine $C^{(2)}$ or $D^{(0)}$ by fitting the {\NNLO} amplitude to the empirical phase shifts.

To explain in detail how the strong correlation between $\VSO{0}$ and $\VSO{2}$ emerges, we start by writing $\delta^{(2)}(k)$ as 
\begin{equation}
    \delta^{(2)}(k) = C^{(2)} \theta_C(k) + D^{(0)} \theta_D(k) + \theta_{2\pi}(k) \, ,
    \label{equ:delta_theta}
\end{equation} 
where
\begin{align}
    \theta_C(k) &\equiv - \frac{\pi}{2} k e^{-2i\delta^{(0)}} \langle \psi^-_k | \VSO{0} | \psi^+_k \rangle \, , \\
    \theta_D(k) &\equiv - \frac{\pi}{2} k e^{-2i\delta^{(0)}} \langle \psi^-_k | \VSO{2} | \psi^+_k \rangle \, , \\
    \theta_{2\pi}(k) &\equiv - \frac{\pi}{2} k e^{-2i\delta^{(0)}} \langle \psi^-_k | V_{2\pi}^{(0)} | \psi^+_k \rangle \, .
\end{align}
The perturbative unitarity~\eqref{equ:delta_rho} ensures that $\theta_C(k)$, $\theta_D(k)$, and $\theta_{2\pi}(k)$ are all real. In addition, $\theta_C(k)$ and $\theta_D(k)$ are entirely decided by the value of $C^{(0)}(\Lambda)$, hence $\delta_\text{In}$. The separable nature of $\VSO{0}$ and $\VSO{2}$ allows us to factorize $\theta_C(k)$ and $\theta_D(k)$ as
\begin{equation}
\begin{aligned}
    \theta_C(k) &= - \frac{\pi}{2} k e^{-2i\delta^{(0)}}\times [\chi(k)]^2
    \, , \\
    \theta_D(k) &= - \frac{\pi}{2} k e^{-2i\delta^{(0)}}\times 2 \chi(k) \phi(k)
    \, .
    \label{equ:theta_psi}
\end{aligned}
\end{equation}

To extract the values of $C^{(2)}$ and $D^{(0)}$, one can use the PWA phase shifts at two different $k$'s as the inputs and set up a group of linear equations to solve for $C^{(2)}$ and $D^{(0)}$:
\begin{equation}
\begin{aligned}
    \delta_\text{PWA}(k_1)- \delta^{(0)}(k_1) &= C^{(2)} \theta_C(k_1) + D^{(0)} \theta_D(k_1) + \theta_{2\pi}(k_1) \, , \\
    \delta_\text{PWA}(k_2) - \delta^{(0)}(k_2) &= C^{(2)} \theta_C(k_2) + D^{(0)} \theta_D(k_2) + \theta_{2\pi}(k_2) \, .
    \label{equ:delta_theta_k1k2}    
\end{aligned}
\end{equation} 

However, if $\theta_C(k)$ and $\theta_D(k)$ become correlated, i.e., not independent functions of $k$, at some cutoff value $\Lambda$, one will not be able to determine $C^{(2)}$ and $D^{(0)}$ at $\Lambda$. This will jeopardize RG invariance in the sense that the EFT can not be implemented at the said $\Lambda$. We will examine whether $\theta_C(k)$ and $\theta_D(k)$ are independent of each other by studying the following dimensionless quantity:
\begin{equation}
    \widetilde{g}(k; \Lambda, \delta_{\text{In}}) \equiv \Lambda^2\, \frac{\theta_C(k)}{\theta_D(k)} \Bigg{|}^k_0 = \Lambda^2 \frac{\chi(k)}{2\phi(k)} \Bigg{|}^k_0 \, ,
    \label{equ:tilde_gk_deltaIN}
\end{equation}
where we have indicated the implicit dependence of $\widetilde{g}(k)$ on $\delta_\text{In}$ and $\Lambda$. By definition $\widetilde{g}(k)$ is always zero at $k = 0$. 

If $\tilde{g}(k; \Lambda)$ vanishes identically in the entire ChEFT kinematic domain of $k$, one will not be able to determine $C^{(2)}$ and $D^{(0)}$ at all from the data. But this does not happen except for $\Lambda_\star$, which, however, we will explain does not pose a conceptual threat, as also argued in Ref.~\cite{Gasparyan:2022isg}. At $\Lambda_\star$ where $C^{(0)}(\Lambda)$ diverges, $\widetilde{g}(k; \Lambda_\star)$ equals zero because although $\chi(k)$ vanishes $\phi(k)$ does not have to, as suggested by Eqs.~\eqref{eqn:ChiKLambdaStar} and \eqref{eqn:DefChi2K}. Equation~\eqref{eqn:ChiKLambdaStar} shows that the diminishing part of $\chi(k; \Lambda)$ near $\Lambda_\star$ is factorized out of its $k$ dependence; therefore, one can let $C^{(2)}(\Lambda)$ and $D^{(0)}(\Lambda)$ diverge as fast as $\theta_C(k; \Lambda)$ and $\theta_D(k; \Lambda)$ approach zeros,
\begin{equation}
    C^{(2)}(\Lambda) \propto (\Lambda - \Lambda_\star)^{-2\alpha} \, , \quad
    D^{(0)}(\Lambda) \propto (\Lambda - \Lambda_\star)^{-\alpha} \, ,
\end{equation}
so that $C^{(2)}(\Lambda) \theta_C(k; \Lambda)$ and $D^{(0)}(\Lambda) \theta_C(k; \Lambda)$ are finite and independent functions of $k$. It follows that even though $\tilde{g}(k; \Lambda) \to 0$ as $\Lambda \to \Lambda_\star$, $\theta_C(k)$ and $\theta_D(k)$ are not correlated.

There are two detrimental scenarios. In the first one, $\tilde{g}(k)$ has identical values at two different $k$'s: $\tilde{g}(k_1) = \tilde{g}(k_2)$. Therefore, one can not use the phase shifts at $k_1$ and $k_2$ to solve for $C^{(2)}$ and $D^{(0)}$ since the two equations in Eq.~\eqref{equ:delta_theta_k1k2} will be linearly dependent. One has to exclude the combination of $k_1$ and $k_2$ when deciding which PWA phase shifts to be used. Although this causes inconvenience in extracting LECs from the data, it does not doom the EFT. In fact a similar situation is reported in Ref.~\cite{Long:2011xw}: The mixing angle $\epsilon_1$, $S$, and $D$-wave phase shifts of $\csd$ at the same $k$ can not form an independent set of constraints to determine the three {\NNLO} contact potentials.

\begin{figure}
    \centering
    \includegraphics[width = 0.4\textwidth]{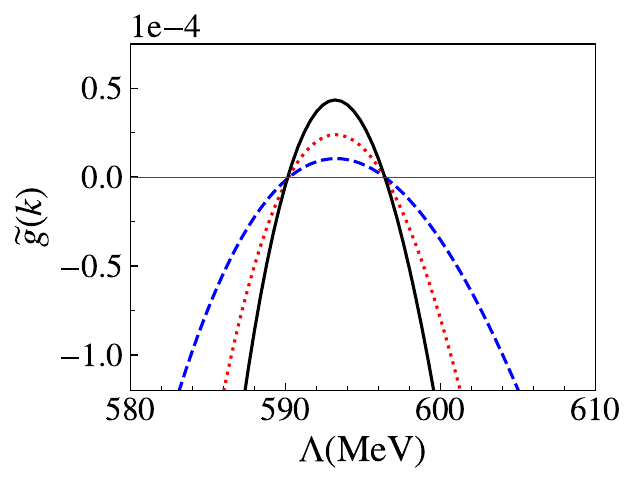}
    \includegraphics[width = 0.4\textwidth]{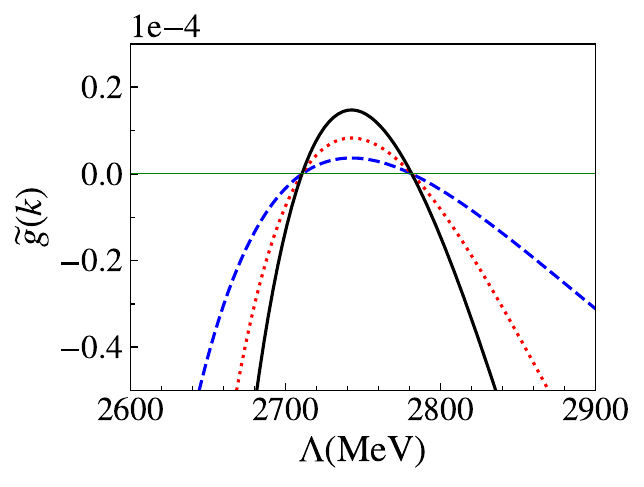}
    \caption{$\widetilde{g}(k; \Lambda)$ as functions of $\Lambda$ at various values of $k$.
    The dashed, dotted, and solid curves correspond to $k=100/150/200$ MeV, respectively. 
    } 
    \label{fig:gk_L}
\end{figure}

The second scenario is more harmful. If $\tilde{g}(k)$ becomes sufficiently small in the ChEFT domain in special cutoff windows, one will not be able to extract values of $C^{(2)}$ and $D^{(0)}$ either, due to the limited precision of almost inevitable numerical calculations in nuclear physics. We plot $\widetilde{g}(k; \Lambda)$ against $\Lambda$ in Fig.~\ref{fig:gk_L} with several representative values of $k$. The curves indicate that $\widetilde{g}(k; \Lambda)$ has four zeros for $\Lambda \leqslant 3200$ MeV (only their vicinities are shown), while more zeros are expected to be found if higher values of $\Lambda$ are investigated. The zeros appear in pairs: One in the neighborhood of $595$ MeV and the other $2750$ MeV. Of each pair, the one with larger $\Lambda$ turns out to be one of $\Lambda_\star$'s, and the other is what really concerns us in the paper, referred to in Ref.~\cite{Gasparyan:2022isg} as genuine exceptional cutoffs and denoted by $\Lambda_E$ in this paper. While $\tilde{g}(k; \Lambda_\star, \delta_\text{In})$ vanishes universally at all finite $k$'s for a given $\delta_\text{In}$, the value of $\Lambda_E$, defined by $\tilde{g}(k; \Lambda_E(k, \delta_\text{In}), \delta_\text{In}) = 0$, varies very slowly as a function of $k$. For instance, $\Delta \Lambda_E \equiv |\Lambda_E(300 \text{MeV}) - \Lambda_E(0)| \approx 0.15$ MeV near $\Lambda = 2710$ MeV. The size of $\Delta \Lambda_E$ sets the goal for the numerical precision of $\Lambda_E$. In order to achieve this goal, we find it necessary to specify $\delta_{\text{In}}$ in degrees to the third decimal place. In plotting Fig.~\ref{fig:gk_L} we have used $\delta_{\text{In}} = \delta_{\text{PWA}} = 10.698^\circ$. But this does not mean that the EFT truncation error must be as small.

\begin{figure}
    \centering
    \includegraphics[width = 0.4\textwidth]{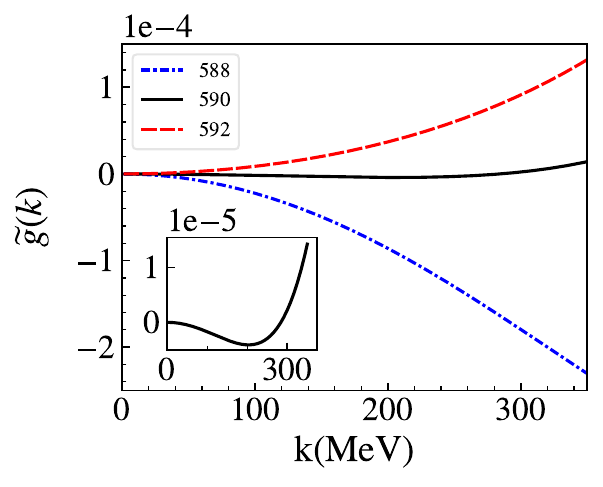}
    \includegraphics[width = 0.4\textwidth]{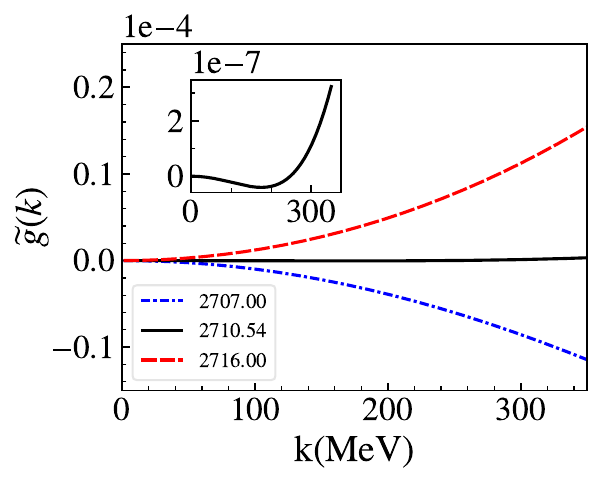}
    \caption{$\widetilde{g}(k; \Lambda)$ as a function of $k$ for various values of $\Lambda$. The numbers in the legends are the values of $\Lambda$ in the unit of MeV.
    }
    \label{fig:gk_fig_ori}
\end{figure}

$\widetilde{g}(k, \Lambda)$ nearly vanishing in the proximity of these two genuine exceptional cutoffs is better demonstrated as a function of $k$ in Fig.~\ref{fig:gk_fig_ori}. When the cutoff value varies from below to above $\Lambda_E$, $\widetilde{g}(k; \Lambda)$ could become arbitrarily small for $k \lesssim 300$ MeV. For instance, $\widetilde{g}(k; \Lambda = 590 \text{MeV}) \sim 10^{-6}$ and $\widetilde{g}(k; \Lambda = 2710.54 \text{MeV}) \sim 10^{-7}$, as illustrated by the inset curves in Fig.~\ref{fig:gk_fig_ori}. Consistent with the observation that $\Lambda_E(k, \delta_\text{In})$ changes slightly with $k$, $\widetilde{g}(k, \Lambda)$ does not vanish uniformly with $k$. Therefore one could in principle pick a set of $k_i$ with distinct values of $\tilde{g}(k_i)$ so that the PWA phase shifts $\delta_\text{PWA}(k_i)$ provide independent inputs. For example, neither $\widetilde{g}(k; \Lambda = 590 \text{MeV})$ nor $\widetilde{g}(k; \Lambda = 2710.54 \text{MeV})$ in Fig.~\ref{fig:gk_fig_ori} is monotonic, so one can use, say, the PWA phase shifts at $k$ that is to the left of the minimum point of $\tilde{g}(k)$.
However, the small magnitude of $\tilde{g}(k)$ at $\Lambda_E$ makes it extremely difficult to numerically solve the equations~\eqref{equ:delta_theta_k1k2} or fit to the PWA phase shifts because the constraints are approximately dependent. In Sec.~\ref{sec:NNLO}, we will show the ``poor'' fits obtained in double-precision calculations. 

In Figs.~\ref{fig:gk_L} and \ref{fig:gk_fig_ori}, we have used an identical value of $\delta_\text{In}$ for all $\Lambda$'s, but this is not necessary, as argued towards the end of Sec.~\ref{sec:LO}. In fact, if we choose a slightly different value for 
$\delta_{\text{In}}$ in the neighborhood of $\Lambda_E$
we can move the zeros away so that $\widetilde{g}(k; \Lambda)$ is no longer arbitrarily small near $\Lambda_E$. For definiteness, we choose to make 
$\delta_{\text{In}}$ vary with $\Lambda$ as follows:
\begin{equation}
    \delta_{\text{In}}(\Lambda) = \left\{
    \begin{aligned}
        &\delta_{\text{PWA}} + \Delta \delta_\text{In}\, ,
        \quad \Lambda \in[588,592]\cup[2707,2716]\, , \\
        &\delta_{\text{PWA}} \, , \quad \text{the rest} \, .
    \end{aligned}
    \right.
    \label{equ:delta_In_L}
\end{equation}
$\Delta \delta_\text{In}$ can be as large as the EFT truncation error, in this case, the {\NNLO} correction to the $\chp{3}{0}$ phase shift, $\delta_{\text{PWA}} (\frac{\mu_0}{M_\text{hi}})^2 \approx 2^\circ$ where we have chosen the break down scale $M_\text{hi}$ to be the nucleon-delta mass splitting. For now, we choose $\Delta \delta_\text{In} = 0.5^\circ$.

\begin{figure}
    \centering
    \includegraphics[width = 0.40\textwidth]{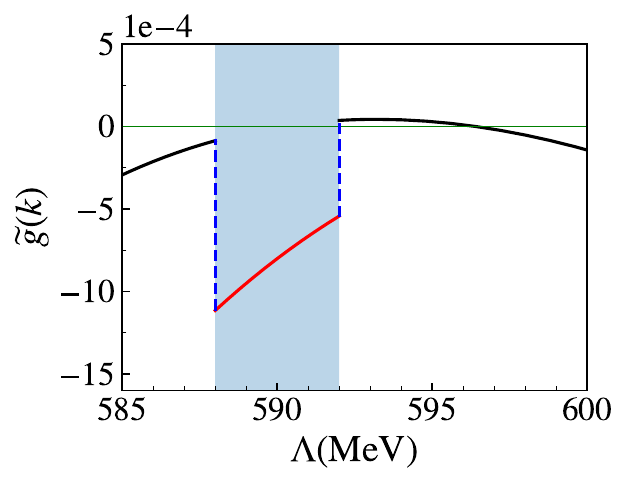}
    \includegraphics[width = 0.40\textwidth]{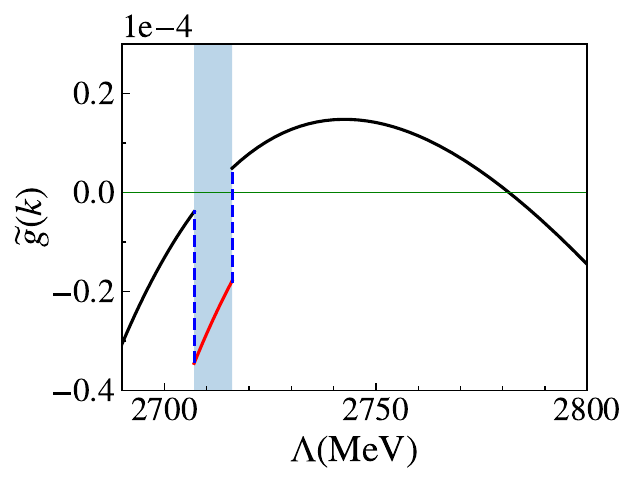}
    \caption{
    $\widetilde{g}(k; \Lambda)$ as a function of $\Lambda$ at $k = 200$ MeV. The bands indicate the intervals of $\Lambda$ where $\delta_{\text{In}}$ takes an alternative value as defined in Eq.~\eqref{equ:delta_In_L}.
    }
    \label{fig:gk_L_fill}
\end{figure}

\begin{figure}
    \centering
    \includegraphics[width = 0.4\textwidth]{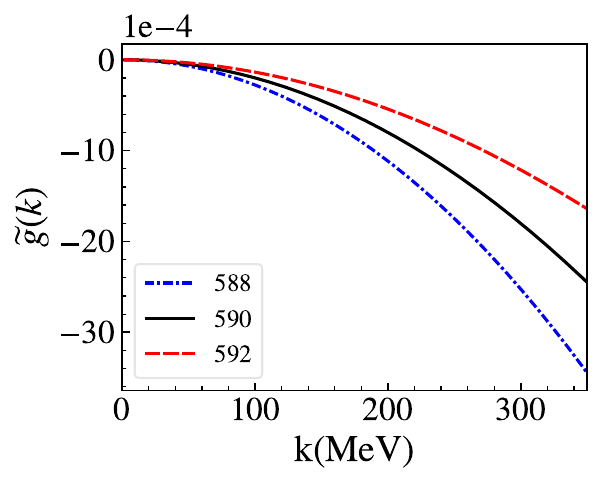}
    \includegraphics[width = 0.4\textwidth]{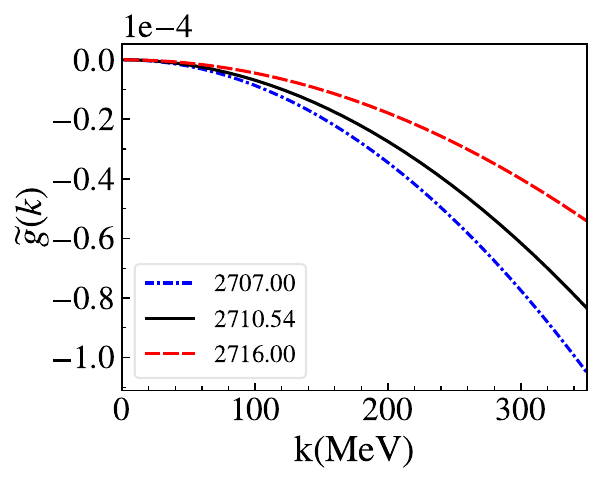}
    \caption{$\widetilde{g}(k;\Lambda)$ as a function of $k$ for several values of $\Lambda$ when $\delta_{\text{In}}$ is varied with $\Lambda$ as prescribed by Eq.~\eqref{equ:delta_In_L}. The numbers in the legends are the values of $\Lambda$ in the unit of MeV.}
    \label{fig:gk_fig_new}
\end{figure}

Plotted in Fig.~\ref{fig:gk_L_fill}, $\tilde{g}(k; \Lambda)$ is shown to have avoided crossing the horizontal line of $\tilde{g} = 0$ near $\Lambda = 590$ MeV and $2710$ MeV. Consequently, $\tilde{g}(k)$ no longer becomes vanishingly small in the ChEFT kinematic domain. In Fig.~\ref{fig:gk_fig_new}, $\tilde{g}(k; \Lambda)$ is shown as a function of $k$, with $\Lambda$ varied within the two $\Lambda_E$ windows where $\delta_{\text{In}}$ takes the alternative value. One finds that $\tilde{g}(k; \Lambda)$ does not cross the $k$ axis as $\Lambda$ varies, unlike in Fig.~\ref{fig:gk_fig_ori}, and the correlation between $\theta_C(k)$ and $\theta_D(k)$ is thus removed in these cutoff windows. We will see in next section that once the independence of $\theta_C(k)$ from $\theta_D(k)$ is achieved, the {\NNLO} contact forces are able to remove those discontinuous cutoff variations.

\section{RG invariance at {\NNLO}\label{sec:NNLO}}

To determine $C^{(2)}$ and $D^{(0)}$, one can use the empirical phase shifts at two different $k$'s as the inputs and solve for the values of $C^{(2)}$ and $D^{(0)}$ using Eq.~\eqref{equ:delta_theta_k1k2}. This is what was done in Refs.~\cite{Long:2007vp, Long:2011qx, Long:2011xw}. In practice, however, it is more common to fit the EFT amplitudes to PWA phase shifts or the scattering data if there are more than one LEC. In this paper we determine the values of $C^{(2)}$ and $D^{(0)}$ by a least-squares fit to the Nijmegen phase shifts, i.e., by minimizing the following function:
\begin{equation}
    R^2(C^{(2)}, D^{(0)})=\sum_i^n[\delta_\text{EFT}(k_i)-\delta_\text{PWA}(k_i)]^2 \, ,    
\end{equation}
where each phase shift $\delta_\text{PWA}(k_i)$ is given equal weight. Phase shift points up to $k = 300$ MeV are used in our fits, which will lead to  better agreement with the PWA than the {\NNLO} results shown in Ref.~\cite{Long:2011qx} where only two phase-shift points were used. 

\begin{figure}
    \centering
    \includegraphics[width = 0.4\textwidth]{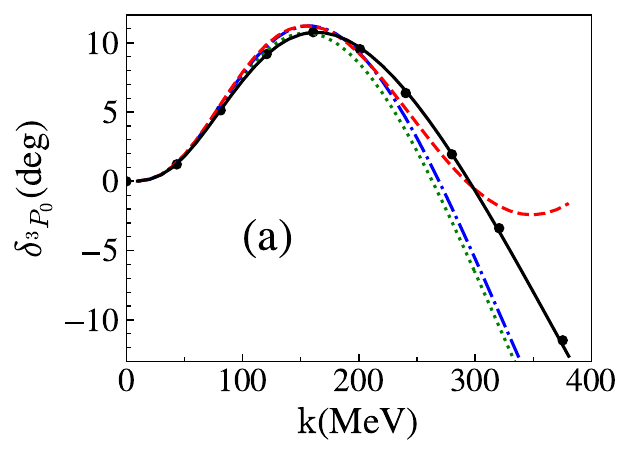}
    \includegraphics[width = 0.4\textwidth]{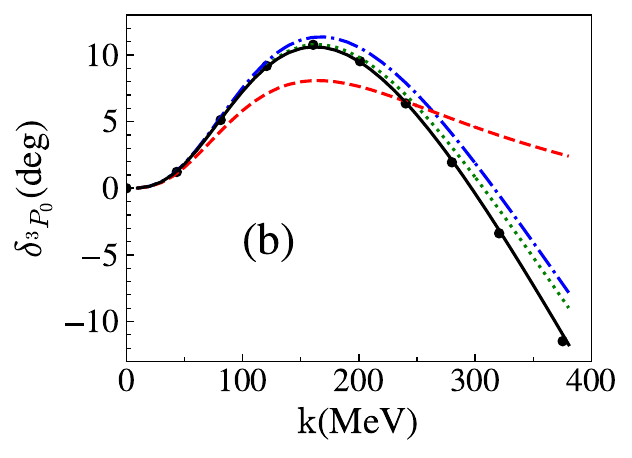}
    \caption{The $\chp{3}{0}$ phase shifts as a function of $k$ for $\Lambda = 590$ MeV (a) and for $\Lambda = 2710.54$ MeV (b). The dash-dotted (LO) and the solid ({\NNLO}) curves use an alternative $\delta_\text{In}$ for the LO input while the dotted (LO) and dashed ({\NNLO}) curves use the unchanged $\delta_\text{In}$.
    The solid dots are from the PWA.}
    \label{fig:phase_2710}
\end{figure}

We first show that in the two exceptional cutoff windows the alternative $\delta_{\text{In}}$ enables the {\NNLO} EFT amplitude to describe the PWA phase shifts. In Fig.~\ref{fig:phase_2710} where the EFT phase shifts are plotted with $\Lambda = 590$ and $2710.54$ MeV, an excellent agreement for $k \lesssim 300$ MeV by the {\NNLO} amplitude is reached. The results are obtained with double-precision calculations. The {\NNLO} phase shifts with unchanged $\delta_\text{In}$ are also shown for comparison. We notice that because $\widetilde{g}(k; \Lambda = 590 \text{MeV})$ is larger than $\widetilde{g}(k; \Lambda = 2710.54 \text{MeV})$ by two orders of magnitude (cf. Fig.~\ref{fig:gk_fig_ori}), the fixed-$\delta_\text{In}$ scheme fares better at $\Lambda = 590$ MeV. Although the two schemes yield similar LO phase shifts, the fixed-$\delta_\text{In}$ scheme can not achieve even qualitative agreement with the PWA towards higher momenta. Therefore, not only $\theta_C(k)$ and $\theta_D(k)$ are now made independent of each other but with suitable $C^{(2)}$ and $D^{(0)}$ they are also able to describe the phase shifts. 

\begin{figure}
    \centering
    \includegraphics[width = 0.4\textwidth]{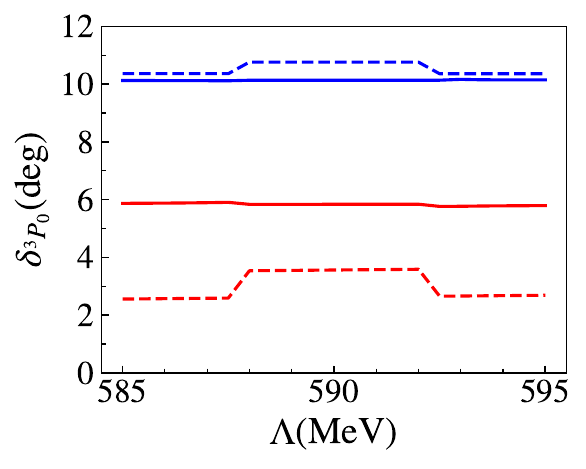}
    \includegraphics[width = 0.4\textwidth]{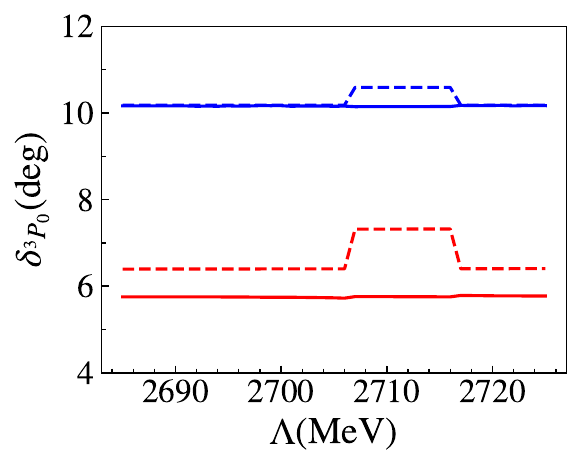}
    \caption{The $\chp{3}{0}$ phase shifts as a function of $\Lambda$ at a given $k$.
    The dashed and solid curves correspond to the LO and {\NNLO} phase shifts, respectively. Different colors indicate the value of $k$ used: $k=137$ MeV (blue) and $k=247$ MeV (red).
    }
    \label{fig:pvl_mixed}
\end{figure}

We still need to examine RG invariance of the amplitudes with the strategy of varying $\delta_\text{In}$. The LO and {\NNLO} phase shifts at $k = 137$ and $247$ MeV are plotted in Fig.~\ref{fig:pvl_mixed}, near and within the two $\Lambda_E$ windows. Shifting $\delta_{\text{In}}$ near $\Lambda_E$, as specified in Eq.~\eqref{equ:delta_In_L}, will no doubt increase the cutoff variation of the LO in those exceptional cutoff windows, shown as the bumps of the dashed curves. 
At {\NNLO}, however, the bumps are removed once $C^{(2)}$ and $D^{(0)}$ retain the ability to contribute independently to $T^{(2)}$ so that $T^{(2)}$ is able to cancel the LO cutoff variation and $T^{(0)} + T^{(2)}$ reaches a higher-level RG invariance. 

\begin{figure}
    \centering
    \includegraphics[width = 0.5\textwidth]{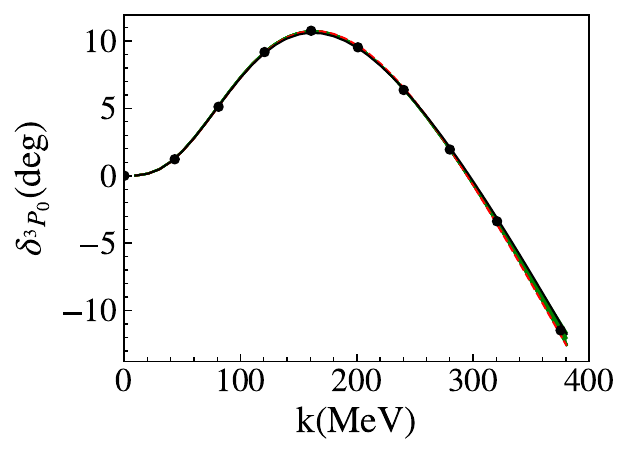}
    \caption{The {\NNLO} $\chp{3}{0}$ phase shifts as a function of $k$. The green band is produced by varying $\Lambda$ from $800$ MeV to $3200$ MeV with an unchanged $\delta_\text{In}$ and the exceptional cutoff windows are ignored. The red dashed and the black solid lines correspond to the {\NNLO} results for $\Lambda = 590$ MeV and for $\Lambda = 2710.54$ MeV, respectively, where the alternative $\delta_\text{In}$ was used. The solid dots are from the PWA.}
    \label{fig:phase_NNLO_band}
\end{figure}

In Fig.~\ref{fig:phase_NNLO_band} we show the overall cutoff variation of the {\NNLO} phase shifts as a function of $k$. The plot is made up of a band--- the {\NNLO} phase shifts for normal values of $\Lambda$ between 800 and 3200 MeV, defined as part of ``the rest'' in Eq.~\eqref{equ:delta_In_L}--- and the curves for the two representative exceptional cutoffs--- the {\NNLO} phase shifts shown in the panels of Fig.~\ref{fig:phase_2710}. The cutoff sensitivity is so small that the band and curves are indistinguishable in the scale of the plot.

\section{Summary and Conclusion\label{sec:summary}}

Renormalization-group invariance has been a valuable guiding principle for building ChEFT-based nuclear forces. It is particularly useful in assessing \textit{a priori} importance of short-range forces that are not constrained otherwise, e.g., by symmetries, with some of the applications referred to in Sec.~\ref{sec:intro}. Renormalization of nonperturbative OPE turns out to be quite complicated, especially in partial waves where OPE is singular and attractive. As a matter of fact, the first $\chp{3}{0}$ counterterm $\VSO{0}$ that has two powers of momenta and is considered {\NNLO} in naive dimensional analysis, was promoted to LO in order to achieve RG invariance~\cite{Nogga:2005hy}.

Another guideline that we adhere to in testing a power counting is to treat subleading terms in perturbation theory, as opposed to iterating them nonperturbatively together with LO. This prevents subleading counterterms from playing more important roles in the intermediate states than they should. For many-body systems where perturbative calculations can be quite difficult, a nonperturbative alternative could be developed \emph{after} the power counting is tested in the few-body sector where one can afford perturbative calculations.

Combining RG invariance and perturbation theory for subleading orders, Ref.~\cite{Long:2011qx} concluded that the $Q^4$ $\chp{3}{0}$ counterterm $\VSO{2}$ must appear at {\NNLO}. In both Ref.~\cite{Long:2011qx} and the present paper, the modified power counting showed a rapid convergence in EFT expansion, in comparison with an implementation of the Weinberg scheme in Ref.\cite{Reinert:2017usi} (cf. Fig.~8). But there is an issue with this perturbative formulation of subleading contact terms, as reported in Ref.~\cite{Gasparyan:2022isg}, which we resolve in the present paper. The matrix elements of the {\NNLO} $\chp{3}{0}$ contact potentials $\VSO{0}$ and $\VSO{2}$ become approximately correlated, hence nearly dependent of each other, at the so-called genuine exceptional cutoffs $\Lambda_E$. As a result, the numerical precision needed to determine subleading LECs can not be achieved by typical double-precision calculations. This difficulty of implementing ChEFT at {\NNLO} for $\Lambda_E$ seemingly violates RG invariance, but it does not mean that one has to abandon RG invariance for a particular EFT of chiral nuclear forces. We showed that if we adjust the strategy of how LECs are determined from the data RG invariance can be retained.

In principle, one can resolve the issue by i) choosing carefully the kinematic points where the PWA phase shifts are taken as the {\NNLO} inputs and ii) resorting to numerical calculations with sufficiently high precision. We propose another solution in the paper that may be more practical in some applications. It centers around a strategy to remove the said approximate correlation between $\VSO{0}$ and $\VSO{2}$. We found that the correlation is intimately related to how the LO LEC $C^{(0)}$ is determined: Those exceptional cutoffs are the consequence of fitting to an identical value of $\delta_\text{In}$ at $k = \mu_0$ for all $\Lambda$'s. This scheme of determining $C^{(0)}$ favors $k = \mu_0$ as it implements perfect cutoff invariance at $k = \mu_0$ while leaving behind residual cutoff dependence everywhere else. Exploiting the EFT tolerance of having uncertainties at each order, we tune $\delta_\text{In}$ by a small deviation $\Delta \delta_\text{In}$, as long as $\Delta \delta_\text{In}$ is on par with the subleading corrections prescribed by the power counting. Similar procedures of slightly modifying LO for various purposes can be found in the EFT literature. For instance, there were applications of pionless EFT where the two-body LO interactions were customized to improve convergence of the EFT expansion in presence of external probes or in many-body systems~\cite{Phillips:1999hh, Contessi:2023yoz}.

We showed that this strategy eliminates the correlation between $\VSO{0}$ and $\VSO{2}$ at $\Lambda_E$, and enables extraction of the {\NNLO} LECs $C^{(2)}$ and $D^{(0)}$ near $\Lambda_E$, making those exceptional cutoff windows amenable to ChEFT. The small variation of the LO phase shifts induced by $\Delta \delta_\text{In}$ near $\Lambda_E$ can be removed at {\NNLO}, therefore improving RG invariance in a way consistent with the power counting.

\acknowledgments

This work was supported by the National Natural Science Foundation of China (NSFC) under Grant Nos. 12275185, 12335002, 12347154, 12335007, 11921006 and 12035001. We thank Jordy de Vries, Sebastian Koenig, Manuel Pavón Valderrama, Bira van Kolck, and Chieh-Jen Yang for useful communications. The High-Performance Computing Platform of Peking University is acknowledged for providing computational resources.

\bibliography{refs.bib}

\begin{thebibliography}{35}
\expandafter\ifx\csname natexlab\endcsname\relax\def\natexlab#1{#1}\fi
\expandafter\ifx\csname bibnamefont\endcsname\relax
  \def\bibnamefont#1{#1}\fi
\expandafter\ifx\csname bibfnamefont\endcsname\relax
  \def\bibfnamefont#1{#1}\fi
\expandafter\ifx\csname citenamefont\endcsname\relax
  \def\citenamefont#1{#1}\fi
\expandafter\ifx\csname url\endcsname\relax
  \def\url#1{\texttt{#1}}\fi
\expandafter\ifx\csname urlprefix\endcsname\relax\def\urlprefix{URL }\fi
\providecommand{\bibinfo}[2]{#2}
\providecommand{\eprint}[2][]{\url{#2}}

\bibitem[{\citenamefont{Gasparyan and Epelbaum}(2023)}]{Gasparyan:2022isg}
\bibinfo{author}{\bibfnamefont{A.~M.} \bibnamefont{Gasparyan}}
  \bibnamefont{and} \bibinfo{author}{\bibfnamefont{E.}~\bibnamefont{Epelbaum}},
  \bibinfo{journal}{Phys. Rev. C} \textbf{\bibinfo{volume}{107}},
  \bibinfo{pages}{034001} (\bibinfo{year}{2023}), \eprint{2210.16225}.

\bibitem[{\citenamefont{Kaplan et~al.}(1996)\citenamefont{Kaplan, Savage, and
  Wise}}]{Kaplan:1996xu}
\bibinfo{author}{\bibfnamefont{D.~B.} \bibnamefont{Kaplan}},
  \bibinfo{author}{\bibfnamefont{M.~J.} \bibnamefont{Savage}},
  \bibnamefont{and} \bibinfo{author}{\bibfnamefont{M.~B.} \bibnamefont{Wise}},
  \bibinfo{journal}{Nucl. Phys. B} \textbf{\bibinfo{volume}{478}},
  \bibinfo{pages}{629} (\bibinfo{year}{1996}), \eprint{nucl-th/9605002}.

\bibitem[{\citenamefont{Nogga et~al.}(2005)\citenamefont{Nogga, Timmermans, and
  van Kolck}}]{Nogga:2005hy}
\bibinfo{author}{\bibfnamefont{A.}~\bibnamefont{Nogga}},
  \bibinfo{author}{\bibfnamefont{R.~G.~E.} \bibnamefont{Timmermans}},
  \bibnamefont{and} \bibinfo{author}{\bibfnamefont{U.}~\bibnamefont{van
  Kolck}}, \bibinfo{journal}{Phys. Rev. C} \textbf{\bibinfo{volume}{72}},
  \bibinfo{pages}{054006} (\bibinfo{year}{2005}), \eprint{nucl-th/0506005}.

\bibitem[{\citenamefont{Birse}(2006)}]{Birse:2005um}
\bibinfo{author}{\bibfnamefont{M.~C.} \bibnamefont{Birse}},
  \bibinfo{journal}{Phys. Rev. C} \textbf{\bibinfo{volume}{74}},
  \bibinfo{pages}{014003} (\bibinfo{year}{2006}), \eprint{nucl-th/0507077}.

\bibitem[{\citenamefont{Long and Yang}(2012{\natexlab{a}})}]{Long:2011xw}
\bibinfo{author}{\bibfnamefont{B.}~\bibnamefont{Long}} \bibnamefont{and}
  \bibinfo{author}{\bibfnamefont{C.~J.} \bibnamefont{Yang}},
  \bibinfo{journal}{Phys. Rev. C} \textbf{\bibinfo{volume}{85}},
  \bibinfo{pages}{034002} (\bibinfo{year}{2012}{\natexlab{a}}),
  \eprint{1111.3993}.

\bibitem[{\citenamefont{Long and Yang}(2012{\natexlab{b}})}]{Long:2012ve}
\bibinfo{author}{\bibfnamefont{B.}~\bibnamefont{Long}} \bibnamefont{and}
  \bibinfo{author}{\bibfnamefont{C.~J.} \bibnamefont{Yang}},
  \bibinfo{journal}{Phys. Rev. C} \textbf{\bibinfo{volume}{86}},
  \bibinfo{pages}{024001} (\bibinfo{year}{2012}{\natexlab{b}}),
  \eprint{1202.4053}.

\bibitem[{\citenamefont{Pav\'on~Valderrama
  et~al.}(2017)\citenamefont{Pav\'on~Valderrama, S\'anchez~S\'anchez, Yang,
  Long, Carbonell, and van Kolck}}]{PavonValderrama:2016lqn}
\bibinfo{author}{\bibfnamefont{M.}~\bibnamefont{Pav\'on~Valderrama}},
  \bibinfo{author}{\bibfnamefont{M.}~\bibnamefont{S\'anchez~S\'anchez}},
  \bibinfo{author}{\bibfnamefont{C.~J.} \bibnamefont{Yang}},
  \bibinfo{author}{\bibfnamefont{B.}~\bibnamefont{Long}},
  \bibinfo{author}{\bibfnamefont{J.}~\bibnamefont{Carbonell}},
  \bibnamefont{and} \bibinfo{author}{\bibfnamefont{U.}~\bibnamefont{van
  Kolck}}, \bibinfo{journal}{Phys. Rev. C} \textbf{\bibinfo{volume}{95}},
  \bibinfo{pages}{054001} (\bibinfo{year}{2017}), \eprint{1611.10175}.

\bibitem[{\citenamefont{Hammer et~al.}(2020)\citenamefont{Hammer, K\"onig, and
  van Kolck}}]{Bira:2020rv}
\bibinfo{author}{\bibfnamefont{H.-W.} \bibnamefont{Hammer}},
  \bibinfo{author}{\bibfnamefont{S.}~\bibnamefont{K\"onig}}, \bibnamefont{and}
  \bibinfo{author}{\bibfnamefont{U.}~\bibnamefont{van Kolck}},
  \bibinfo{journal}{Rev. Mod. Phys.} \textbf{\bibinfo{volume}{92}},
  \bibinfo{pages}{025004} (\bibinfo{year}{2020}).

\bibitem[{\citenamefont{Pav\'on~Valderrama and
  Phillips}(2015)}]{PavonValderrama:2014zeq}
\bibinfo{author}{\bibfnamefont{M.}~\bibnamefont{Pav\'on~Valderrama}}
  \bibnamefont{and} \bibinfo{author}{\bibfnamefont{D.~R.}
  \bibnamefont{Phillips}}, \bibinfo{journal}{Phys. Rev. Lett.}
  \textbf{\bibinfo{volume}{114}}, \bibinfo{pages}{082502}
  (\bibinfo{year}{2015}), \eprint{1407.0437}.

\bibitem[{\citenamefont{Shi et~al.}(2022)\citenamefont{Shi, Peng, Liu, Lyu, and
  Long}}]{Shi:2022blm}
\bibinfo{author}{\bibfnamefont{W.}~\bibnamefont{Shi}},
  \bibinfo{author}{\bibfnamefont{R.}~\bibnamefont{Peng}},
  \bibinfo{author}{\bibfnamefont{T.-X.} \bibnamefont{Liu}},
  \bibinfo{author}{\bibfnamefont{S.}~\bibnamefont{Lyu}}, \bibnamefont{and}
  \bibinfo{author}{\bibfnamefont{B.}~\bibnamefont{Long}},
  \bibinfo{journal}{Phys. Rev. C} \textbf{\bibinfo{volume}{106}},
  \bibinfo{pages}{015505} (\bibinfo{year}{2022}), \eprint{2205.02000}.

\bibitem[{\citenamefont{Liu et~al.}(2022)\citenamefont{Liu, Peng, Lyu, and
  Long}}]{Liu:2022cfd}
\bibinfo{author}{\bibfnamefont{T.-X.} \bibnamefont{Liu}},
  \bibinfo{author}{\bibfnamefont{R.}~\bibnamefont{Peng}},
  \bibinfo{author}{\bibfnamefont{S.}~\bibnamefont{Lyu}}, \bibnamefont{and}
  \bibinfo{author}{\bibfnamefont{B.}~\bibnamefont{Long}},
  \bibinfo{journal}{Phys. Rev. C} \textbf{\bibinfo{volume}{106}},
  \bibinfo{pages}{055501} (\bibinfo{year}{2022}), \eprint{2207.04241}.

\bibitem[{\citenamefont{de~Vries et~al.}(2021)\citenamefont{de~Vries, Gnech,
  and Shain}}]{deVries:2020loy}
\bibinfo{author}{\bibfnamefont{J.}~\bibnamefont{de~Vries}},
  \bibinfo{author}{\bibfnamefont{A.}~\bibnamefont{Gnech}}, \bibnamefont{and}
  \bibinfo{author}{\bibfnamefont{S.}~\bibnamefont{Shain}},
  \bibinfo{journal}{Phys. Rev. C} \textbf{\bibinfo{volume}{103}},
  \bibinfo{pages}{L012501} (\bibinfo{year}{2021}), \eprint{2007.04927}.

\bibitem[{\citenamefont{Cirigliano et~al.}(2018)\citenamefont{Cirigliano,
  Dekens, De~Vries, Graesser, Mereghetti, Pastore, and
  Van~Kolck}}]{Cirigliano:2018hja}
\bibinfo{author}{\bibfnamefont{V.}~\bibnamefont{Cirigliano}},
  \bibinfo{author}{\bibfnamefont{W.}~\bibnamefont{Dekens}},
  \bibinfo{author}{\bibfnamefont{J.}~\bibnamefont{De~Vries}},
  \bibinfo{author}{\bibfnamefont{M.~L.} \bibnamefont{Graesser}},
  \bibinfo{author}{\bibfnamefont{E.}~\bibnamefont{Mereghetti}},
  \bibinfo{author}{\bibfnamefont{S.}~\bibnamefont{Pastore}}, \bibnamefont{and}
  \bibinfo{author}{\bibfnamefont{U.}~\bibnamefont{Van~Kolck}},
  \bibinfo{journal}{Phys. Rev. Lett.} \textbf{\bibinfo{volume}{120}},
  \bibinfo{pages}{202001} (\bibinfo{year}{2018}), \eprint{1802.10097}.

\bibitem[{\citenamefont{Cirigliano et~al.}(2021)\citenamefont{Cirigliano,
  Dekens, de~Vries, Hoferichter, and Mereghetti}}]{Cirigliano:2020dmx}
\bibinfo{author}{\bibfnamefont{V.}~\bibnamefont{Cirigliano}},
  \bibinfo{author}{\bibfnamefont{W.}~\bibnamefont{Dekens}},
  \bibinfo{author}{\bibfnamefont{J.}~\bibnamefont{de~Vries}},
  \bibinfo{author}{\bibfnamefont{M.}~\bibnamefont{Hoferichter}},
  \bibnamefont{and}
  \bibinfo{author}{\bibfnamefont{E.}~\bibnamefont{Mereghetti}},
  \bibinfo{journal}{Phys. Rev. Lett.} \textbf{\bibinfo{volume}{126}},
  \bibinfo{pages}{172002} (\bibinfo{year}{2021}), \eprint{2012.11602}.

\bibitem[{\citenamefont{FRANK et~al.}(1971)\citenamefont{FRANK, LAND, and
  SPECTOR}}]{RevModPhys.43.36}
\bibinfo{author}{\bibfnamefont{W.~M.} \bibnamefont{FRANK}},
  \bibinfo{author}{\bibfnamefont{D.~J.} \bibnamefont{LAND}}, \bibnamefont{and}
  \bibinfo{author}{\bibfnamefont{R.~M.} \bibnamefont{SPECTOR}},
  \bibinfo{journal}{Rev. Mod. Phys.} \textbf{\bibinfo{volume}{43}},
  \bibinfo{pages}{36} (\bibinfo{year}{1971}).

\bibitem[{\citenamefont{Beane et~al.}(2001)\citenamefont{Beane, Bedaque,
  Childress, Kryjevski, McGuire, and van Kolck}}]{Beane:2000wh}
\bibinfo{author}{\bibfnamefont{S.~R.} \bibnamefont{Beane}},
  \bibinfo{author}{\bibfnamefont{P.~F.} \bibnamefont{Bedaque}},
  \bibinfo{author}{\bibfnamefont{L.}~\bibnamefont{Childress}},
  \bibinfo{author}{\bibfnamefont{A.}~\bibnamefont{Kryjevski}},
  \bibinfo{author}{\bibfnamefont{J.}~\bibnamefont{McGuire}}, \bibnamefont{and}
  \bibinfo{author}{\bibfnamefont{U.}~\bibnamefont{van Kolck}},
  \bibinfo{journal}{Phys. Rev. A} \textbf{\bibinfo{volume}{64}},
  \bibinfo{pages}{042103} (\bibinfo{year}{2001}), \eprint{quant-ph/0010073}.

\bibitem[{\citenamefont{Long and van Kolck}(2008)}]{Long:2007vp}
\bibinfo{author}{\bibfnamefont{B.}~\bibnamefont{Long}} \bibnamefont{and}
  \bibinfo{author}{\bibfnamefont{U.}~\bibnamefont{van Kolck}},
  \bibinfo{journal}{Annals Phys.} \textbf{\bibinfo{volume}{323}},
  \bibinfo{pages}{1304} (\bibinfo{year}{2008}), \eprint{0707.4325}.

\bibitem[{\citenamefont{Hammer and Platter}(2011)}]{Hammer:2011kg}
\bibinfo{author}{\bibfnamefont{H.-W.} \bibnamefont{Hammer}} \bibnamefont{and}
  \bibinfo{author}{\bibfnamefont{L.}~\bibnamefont{Platter}},
  \bibinfo{journal}{Phil. Trans. Roy. Soc. Lond. A}
  \textbf{\bibinfo{volume}{369}}, \bibinfo{pages}{2679} (\bibinfo{year}{2011}),
  \eprint{1102.3789}.

\bibitem[{\citenamefont{Odell et~al.}(2019)\citenamefont{Odell, Deltuva,
  Bonilla, and Platter}}]{Odell:2019wjq}
\bibinfo{author}{\bibfnamefont{D.}~\bibnamefont{Odell}},
  \bibinfo{author}{\bibfnamefont{A.}~\bibnamefont{Deltuva}},
  \bibinfo{author}{\bibfnamefont{J.}~\bibnamefont{Bonilla}}, \bibnamefont{and}
  \bibinfo{author}{\bibfnamefont{L.}~\bibnamefont{Platter}},
  \bibinfo{journal}{Phys. Rev. C} \textbf{\bibinfo{volume}{100}},
  \bibinfo{pages}{054001} (\bibinfo{year}{2019}), \eprint{1903.00034}.

\bibitem[{\citenamefont{Long and Yang}(2011)}]{Long:2011qx}
\bibinfo{author}{\bibfnamefont{B.}~\bibnamefont{Long}} \bibnamefont{and}
  \bibinfo{author}{\bibfnamefont{C.-J.} \bibnamefont{Yang}},
  \bibinfo{journal}{Phys. Rev. C} \textbf{\bibinfo{volume}{84}},
  \bibinfo{pages}{057001} (\bibinfo{year}{2011}).

\bibitem[{\citenamefont{Beane et~al.}(2002)\citenamefont{Beane, Bedaque,
  Savage, and van Kolck}}]{Beane:2001bc}
\bibinfo{author}{\bibfnamefont{S.~R.} \bibnamefont{Beane}},
  \bibinfo{author}{\bibfnamefont{P.~F.} \bibnamefont{Bedaque}},
  \bibinfo{author}{\bibfnamefont{M.~J.} \bibnamefont{Savage}},
  \bibnamefont{and} \bibinfo{author}{\bibfnamefont{U.}~\bibnamefont{van
  Kolck}}, \bibinfo{journal}{Nucl. Phys. A} \textbf{\bibinfo{volume}{700}},
  \bibinfo{pages}{377} (\bibinfo{year}{2002}), \eprint{nucl-th/0104030}.

\bibitem[{\citenamefont{Kaiser et~al.}(1997)\citenamefont{Kaiser, Brockmann,
  and Weise}}]{Kaiser:1997mw}
\bibinfo{author}{\bibfnamefont{N.}~\bibnamefont{Kaiser}},
  \bibinfo{author}{\bibfnamefont{R.}~\bibnamefont{Brockmann}},
  \bibnamefont{and} \bibinfo{author}{\bibfnamefont{W.}~\bibnamefont{Weise}},
  \bibinfo{journal}{Nucl. Phys. A} \textbf{\bibinfo{volume}{625}},
  \bibinfo{pages}{758} (\bibinfo{year}{1997}), \eprint{nucl-th/9706045}.

\bibitem[{\citenamefont{Epelbaum et~al.}(2000)\citenamefont{Epelbaum, Gloeckle,
  and Meissner}}]{Epelbaum:1999dj}
\bibinfo{author}{\bibfnamefont{E.}~\bibnamefont{Epelbaum}},
  \bibinfo{author}{\bibfnamefont{W.}~\bibnamefont{Gloeckle}}, \bibnamefont{and}
  \bibinfo{author}{\bibfnamefont{U.-G.} \bibnamefont{Meissner}},
  \bibinfo{journal}{Nucl. Phys. A} \textbf{\bibinfo{volume}{671}},
  \bibinfo{pages}{295} (\bibinfo{year}{2000}), \eprint{nucl-th/9910064}.

\bibitem[{\citenamefont{Ord\'o\~nez et~al.}(1996)\citenamefont{Ord\'o\~nez,
  Ray, and van Kolck}}]{Bira:1996}
\bibinfo{author}{\bibfnamefont{C.}~\bibnamefont{Ord\'o\~nez}},
  \bibinfo{author}{\bibfnamefont{L.}~\bibnamefont{Ray}}, \bibnamefont{and}
  \bibinfo{author}{\bibfnamefont{U.}~\bibnamefont{van Kolck}},
  \bibinfo{journal}{Phys. Rev. C} \textbf{\bibinfo{volume}{53}},
  \bibinfo{pages}{2086} (\bibinfo{year}{1996}).

\bibitem[{\citenamefont{Epelbaoum et~al.}(1999)\citenamefont{Epelbaoum,
  Glöckle, Krüger, and Meißner}}]{EPELBAOUM1999413}
\bibinfo{author}{\bibfnamefont{E.}~\bibnamefont{Epelbaoum}},
  \bibinfo{author}{\bibfnamefont{W.}~\bibnamefont{Glöckle}},
  \bibinfo{author}{\bibfnamefont{A.}~\bibnamefont{Krüger}}, \bibnamefont{and}
  \bibinfo{author}{\bibfnamefont{U.-G.} \bibnamefont{Meißner}},
  \bibinfo{journal}{Nuclear Physics A} \textbf{\bibinfo{volume}{645}},
  \bibinfo{pages}{413} (\bibinfo{year}{1999}), ISSN \bibinfo{issn}{0375-9474}.

\bibitem[{\citenamefont{Peng et~al.}(2020)\citenamefont{Peng, Lyu, and
  Long}}]{Peng_2020}
\bibinfo{author}{\bibfnamefont{R.}~\bibnamefont{Peng}},
  \bibinfo{author}{\bibfnamefont{S.}~\bibnamefont{Lyu}}, \bibnamefont{and}
  \bibinfo{author}{\bibfnamefont{B.}~\bibnamefont{Long}},
  \bibinfo{journal}{Communications in Theoretical Physics}
  \textbf{\bibinfo{volume}{72}}, \bibinfo{pages}{095301}
  (\bibinfo{year}{2020}).

\bibitem[{nn.()}]{nn.online}
\emph{\bibinfo{title}{The nn-online}},
  \bibinfo{howpublished}{\url{http://nn-online.org}}.

\bibitem[{\citenamefont{Stoks et~al.}(1993)\citenamefont{Stoks, Klomp,
  Rentmeester, and de~Swart}}]{Stoks:1993tb}
\bibinfo{author}{\bibfnamefont{V.~G.~J.} \bibnamefont{Stoks}},
  \bibinfo{author}{\bibfnamefont{R.~A.~M.} \bibnamefont{Klomp}},
  \bibinfo{author}{\bibfnamefont{M.~C.~M.} \bibnamefont{Rentmeester}},
  \bibnamefont{and} \bibinfo{author}{\bibfnamefont{J.~J.}
  \bibnamefont{de~Swart}}, \bibinfo{journal}{Phys. Rev. C}
  \textbf{\bibinfo{volume}{48}}, \bibinfo{pages}{792} (\bibinfo{year}{1993}).

\bibitem[{\citenamefont{Entem and Machleidt}(2002)}]{Entem:2002sf}
\bibinfo{author}{\bibfnamefont{D.~R.} \bibnamefont{Entem}} \bibnamefont{and}
  \bibinfo{author}{\bibfnamefont{R.}~\bibnamefont{Machleidt}},
  \bibinfo{journal}{Phys. Rev. C} \textbf{\bibinfo{volume}{66}},
  \bibinfo{pages}{014002} (\bibinfo{year}{2002}), \eprint{nucl-th/0202039}.

\bibitem[{\citenamefont{van Kolck}(1999)}]{vanKolck:1998bw}
\bibinfo{author}{\bibfnamefont{U.}~\bibnamefont{van Kolck}},
  \bibinfo{journal}{Nucl. Phys. A} \textbf{\bibinfo{volume}{645}},
  \bibinfo{pages}{273} (\bibinfo{year}{1999}), \eprint{nucl-th/9808007}.

\bibitem[{\citenamefont{Pav\'on~Valderrama}(2011{\natexlab{a}})}]{Valderrama:2009ei}
\bibinfo{author}{\bibfnamefont{M.}~\bibnamefont{Pav\'on~Valderrama}},
  \bibinfo{journal}{Phys. Rev. C} \textbf{\bibinfo{volume}{83}},
  \bibinfo{pages}{024003} (\bibinfo{year}{2011}{\natexlab{a}}),
  \eprint{0912.0699}.

\bibitem[{\citenamefont{Pav\'on~Valderrama}(2011{\natexlab{b}})}]{PavonValderrama:2011fcz}
\bibinfo{author}{\bibfnamefont{M.}~\bibnamefont{Pav\'on~Valderrama}},
  \bibinfo{journal}{Phys. Rev. C} \textbf{\bibinfo{volume}{84}},
  \bibinfo{pages}{064002} (\bibinfo{year}{2011}{\natexlab{b}}),
  \eprint{1108.0872}.

\bibitem[{\citenamefont{Reinert et~al.}(2018)\citenamefont{Reinert, Krebs, and
  Epelbaum}}]{Reinert:2017usi}
\bibinfo{author}{\bibfnamefont{P.}~\bibnamefont{Reinert}},
  \bibinfo{author}{\bibfnamefont{H.}~\bibnamefont{Krebs}}, \bibnamefont{and}
  \bibinfo{author}{\bibfnamefont{E.}~\bibnamefont{Epelbaum}},
  \bibinfo{journal}{Eur. Phys. J. A} \textbf{\bibinfo{volume}{54}},
  \bibinfo{pages}{86} (\bibinfo{year}{2018}), \eprint{1711.08821}.

\bibitem[{\citenamefont{Phillips et~al.}(2000)\citenamefont{Phillips, Rupak,
  and Savage}}]{Phillips:1999hh}
\bibinfo{author}{\bibfnamefont{D.~R.} \bibnamefont{Phillips}},
  \bibinfo{author}{\bibfnamefont{G.}~\bibnamefont{Rupak}}, \bibnamefont{and}
  \bibinfo{author}{\bibfnamefont{M.~J.} \bibnamefont{Savage}},
  \bibinfo{journal}{Phys. Lett. B} \textbf{\bibinfo{volume}{473}},
  \bibinfo{pages}{209} (\bibinfo{year}{2000}), \eprint{nucl-th/9908054}.

\bibitem[{\citenamefont{Contessi et~al.}(2024)\citenamefont{Contessi,
  Sch\"afer, and van Kolck}}]{Contessi:2023yoz}
\bibinfo{author}{\bibfnamefont{L.}~\bibnamefont{Contessi}},
  \bibinfo{author}{\bibfnamefont{M.}~\bibnamefont{Sch\"afer}},
  \bibnamefont{and} \bibinfo{author}{\bibfnamefont{U.}~\bibnamefont{van
  Kolck}}, \bibinfo{journal}{Phys. Rev. A} \textbf{\bibinfo{volume}{109}},
  \bibinfo{pages}{022814} (\bibinfo{year}{2024}), \eprint{2310.15760}.

\end{thebibliography}

\end{document}